\documentclass[aps,showpacs,floatfix,preprintnumbers]{revtex4}

\usepackage{amssymb}
\usepackage{graphicx}
\usepackage{dcolumn}% Align table columns on decimal point
\usepackage{bm}% bold math
\usepackage{amsmath}
\usepackage[usenames]{color}

\usepackage{graphicx, epsfig}
\usepackage{color}
\usepackage{amsmath}
\usepackage{dcolumn}% Align table columns on decimal point
\newcommand{\be}{\begin{equation}}
\newcommand{\ee}{\end{equation}}
\newcommand{\bea}{\begin{eqnarray}}
\newcommand{\eea}{\end{eqnarray}}

\newcommand{\gapp}{\mathrel{\raise.3ex\hbox{$>$}\mkern-14mu \lower0.6ex\hbox{$\sim$}}}
\newcommand{\lapp}{\mathrel{\raise.3ex\hbox{$<$}\mkern-14mu \lower0.6ex\hbox{$\sim$}}}
\def\bbox{{\,\lower0.9pt\vbox{\hrule \hbox{\vrule height 0.2 cm
\hskip 0.2 cm \vrule  height 0.2 cm}\hrule}\,}}

\begin{document}
\title{Volume Renormalization and the Higgs}
\author{De-Chang Dai$^1$}
\author{Dejan Stojkovic$^2$}
\affiliation{$^1$ Institute of Natural Sciences, Shanghai Key Lab for Particle Physics and Cosmology, and Center for Astrophysics and Astronomy, Department of Physics and Astronomy, Shanghai Jiao Tong University, Shanghai 200240, China\\ $^2$ HEPCOS, Department of Physics, SUNY at Buffalo, Buffalo, NY 14260-1500}
\date{\today}
\begin{abstract}

Traditionally, Quantum Field Theory (QFT) treats particle excitations as point-like objects, which is the source of ubiquitous divergences. We demonstrate that a minimal modification of QFT with finite volume particles may cure QFT of divergences and illuminate the physics behind the mathematical construct of our theories. The method allows for a non-perturbative treatment of the free field and self-interacting theories (though extensions to all interacting field theories might be possible). In particular, non-perturbatively defined mass is finite. When applied to the standard model Higgs mechanism, the method implies that a finite range of parameters allows for creation of a well defined Higgs particle, whose Compton wavelength is larger than its physical size, in the broken symmetry phase (as small oscillations around the vacuum). This has profound consequences for Higgs production at the LHC. The parameter range in which the Higgs excitation with the mass of $125$GeV behaves as a proper particle is very restricted.

\end{abstract}
\maketitle

\section{Introduction}

 The problem of divergences has been following Quantum Field theory (QFT) from its invention. Most of the diagrams describing physical processes are divergent. To deal with this problem, various renormalization methods were invented. Renormalization proved very useful and robust, giving experimentally verifiable results. However, it is still useful to look for other approaches that may offer a different look to the same problem.

In QFT, particles are taken to be strictly point-like. In the language of mathematics, the fields are operator valued distributions and their products are not well-defined. We will therefore choose a particle volume as a free parameter and study the finite volume quantization procedure. We will concentrate on two cases: (a) free scalar field and (b) the Higgs mechanism. We will derive unexpected results that will make us revisit some basic assumptions used in dealing with physical processes in QFT. For example we use non-perturbative definition of mass in order to extract more physical information than perturbative methods can give us. As a consequence, we will see that there exists a finite range of parameters which allows for production of a well defined Higgs particle in accelerators.

In this paper we will review the role of particle size in the QFT quantization procedure to develop a non-perturbative treatment of free and self-interacting field theories. Our approach is in line with the spirit of the Wilsonian approach; though we effectively introduce a more natural particle volume cutoff instead of Wilsonian cutoff at an {\it a priory} unknown energy.

\section{Second Quantization}

Consider the Lagrangian for a free scalar field, $\phi(x)$.
\begin{equation}
S=\int d^4x \left[ \frac{1}{2}(\partial \phi)^2 -\frac{1}{2} m^2\phi^2\right]
\end{equation}

The equation of motion is
\begin{equation}
\partial_t^2 \phi -\partial_x^2 \phi +m^2 \phi =0
\end{equation}

In QFT we quantize the scalar field  $\phi(x)$ by introducing the canonical momentum with the following commutation relations
%\begin{eqnarray}
\begin{equation}
\label{uncertain}
%&&
[\phi (x),\pi (y)]=i\delta^{(3)}(x-y), \,
%&&
 [\phi (x),\phi (y)]=[\pi (x), \pi (y) ] =0
\end{equation}
%\end{eqnarray}

According to Eq.~(\ref{uncertain}), $\phi(x)$ and $\phi (y)$ are unrelated if $x \neq y$. I.e. excitations of $\phi(x)$ are point particles. In order to construct a local point particle one has to include an integral over all possible momenta
\be
  \phi(x) = \int \frac{d^4k}{(2\pi)^4} e^{ikx} \tilde{\phi}(k)
\ee
which will be the source of ultra-violet divergences once we include these momenta into physical processes \cite{Wilczek:1998ma}. In other words, the Dirac delta function has zero uncertainty in $x$-space, so it must have infinite uncertainty in the momentum space. This infinite momentum causes the divergence in the calculations of Feynman diagrams and it is the primary reason to introduce renormalization.

We adopt here a different quantization approach. Since perhaps there are no truly point particles in the physical world, the commutation relations must be modified. If the wave function of a realistic particle is not a Dirac delta function, then  $\phi(x)$ and $\phi(y)$ must be related and
\begin{equation}
\left[\phi (x), \pi (y)\right]\neq 0
\end{equation}
even if $x\neq y$. We note that this reasoning should be true in the Wilsonian approach if we Fourier transform the fields back after we impose a cut-off in energy, but working in configuration space is more transparent. Setting $x=y$, we generalize the equal time commutation relation to

\begin{equation}
[\phi (x), \pi (x)]= i f(x)^2
\end{equation}
where $f(x)$ is the wave-packet of the particle. In principle it is possible to use a precise form of $f(x)$ \cite{p}, however we will for simplicity assume that the particle is a uniform spherical ball in its own rest frame. Therefore
\begin{equation}
\label{commutation}
[\phi(x), V(x)\pi (x)] =i
\end{equation}
where $V(x)$ is the physical volume of the particle. Throughout the paper we will treat the $V(x)$ as a free parameter. In QFT (e.g. Eq.~(\ref{uncertain})), $V(x) =1/\delta ^{3} (0)$, which is infinitely small. We will see that in order to regularize at least some of infinities, $V(x)$ must be finite. As in the standard procedure, we will concentrate on the plane wave solutions $\phi = A e^{-i(Et -\vec{p}\vec{x})}$.

We emphasis that the commutation relation  Eq.~(\ref{commutation}) is Lorentz invariant. Let's consider the rest frame of the particle. If $V$ is the volume in the rest frame of the particle, then boosted volume is $V'=V/\gamma$, where $\gamma$ is the relativistic factor. The canonical momentum $\pi$, which is just $\pi = \partial \phi/\partial_t = -iE\phi$ (in the rest frame) is boosted as $\pi'=\gamma \pi$. Thus, the commutation relation in Eq.~(\ref{commutation}) is Lorentz invariant.  In an arbitrary frame, there will be additional contribution coming from  the $\partial_x$ derivative, but because of the Lorentz invariance, the physics will remain the same.
Another way to see the same thing is to note that the regularized Dirac delta function of zero argument, $\delta(0)$, is actually an inverse volume. Comparing Eq.~(\ref{uncertain}) with Eq.~(\ref{commutation}), we see that we keep all of the properties of the standard canonical quantization, except that we regularized the infinity of $\delta(0)$ in a Lorentz invariant way.

We will perform the quantization procedure in the rest frame of the particle where $\vec{p}=0$ and $E=m$, so that $\phi = A e^{-imt}$.
The time coordinate in the rest frame is the proper time $\tau$, while the spatial derivatives vanish. Thus, for the plane wave solutions we can write $\partial_\tau^2=\partial_t^2-\partial_x^2$. The Hamiltonian density becomes
\begin{equation}
\label{motion1}
\frac{1}{2}(\partial_\tau \phi)^2 +\frac{1}{2}m^2 \phi^2 =H
\end{equation}
where $H$ is the Hamiltonian or energy density for the scalar field in its rest frame. $H$ will therefore carry only information about the particle mass, and not its momentum, which will be important later when we discuss the eigenvalues of $H$. Quantization in the general frame will be related to quantization in the rest frame by the usual relation $E^2=\vec{p}^2+m^2$.

If we compare Eq.~(\ref{commutation}) with the commutation relation in quantum mechanics we see that $V \pi(x)$ plays the role of momentum, while $\phi(x)$ plays the role of the space coordinate in one dimension. One can then apply the functional Schr$\ddot{\rm o}$dinger formalism, i.e. we can define the wave-functional $\Psi(\phi(x))$ and quantize it.
We choose the canonical quantization instead of the second quantization with creation and annihilation operators because we want to find the amplitude of oscillations of the field $\phi(x)$. Since the functional Schr$\ddot{\rm o}$dinger formalism contains all the information that the second quantization does (see e.g. \cite{Vachaspati:2006ki}), using one or another is just a matter of convenience. The generalized momentum that corresponds to the wave-functional $\Psi(\phi(x))$ is $V\pi(x) =-i\partial_{\phi(x)}$. We can now write the functional Schr$\ddot{\rm o}$dinger equation that corresponds to the Hamiltonian in Eq.~(\ref{motion1}) as
\begin{equation}
\label{motion2}
-\frac{1}{2} \partial_{\phi}^2\Psi(\phi)+\frac{1}{2}V^2 m^2 \phi^2\Psi(\phi) =V^2H\Psi(\phi)
\end{equation}
Eq.~(\ref{motion2}) describes a simple harmonic oscillator with mass $V$, frequency $m$, and Hamiltonian $VH=(n+1/2)m$, where $n$ is the quantum number labeling the energy levels. Therefore, the energy difference from the ground to the first excited state is $m$; the rest energy of the particle. If we take the wave-packet of the particle to be a uniform sphere, then $V$ can be understood as the size of the particle. As we pointed out before, $V=1/\delta^{3}(0)=0$ in QFT, which again implies that particles have no physical size.

We can now find the amplitude of oscillations of the field $\phi(x)$ using the analogy with the simple harmonic oscillator
\begin{equation}\label{pp}
A \equiv <\Psi|\phi^2 |\Psi >=\frac{(2n+1)}{V m}
\end{equation}

In QFT, $V\rightarrow 0$, and the value of $A$ becomes infinitely large. Eq.~(\ref{pp}) also represents the two-point correlation function from $x$ to $x$ in QFT; its infinite value indicates that our procedure is consistent with the formalism of QFT in the limit of $V\rightarrow 0$.

The energy of a free particle does not depend on its volume thus infinite amplitude will not affect it. Therefore, there is no problem with free non-interacting particles. However, once we include interaction terms (even self-interactions) which in general depend on the amplitude of oscillations of the field, an infinite amplitude will cause an infinite contribution to potential. It is then clear why the Feynman diagrams are usually divergent.

\section{Higgs mass}

Let us turn our attention to a most important scalar field - the Higgs. Even self-interactions of the Higgs field have profound consequences on Higgs phenomenology. A large oscillation amplitude would cause serious problems for a Higgs with the standard symmetry breaking Mexican hat potential. If the Higgs field is sitting at the bottom of the potential in the broken symmetry phase, a large enough amplitude would kick it over the barrier. This would make the physical Higgs particle (usually defined as small oscillations at the bottom of the potential) ill defined.  In the standard model, the Higgs field has quadratic and quartic self-interactions, so the physical size of the Higgs particle should also depend on the parameters in the potential that determine these self-interactions.  This finite value will cure the divergence described in Eq.~(\ref{pp}) and may save the standard symmetry breaking picture. However, to verify whether this indeed happens, we have to perform explicit calculations.

After the electroweak symmetry breaking, the Higgs action in unitary gauge is \cite{ps}
\begin{equation} \label{S1}
S=\int d^4x \left[ \frac{(\partial \eta) ^2}{2} - \frac{m_H^2 \eta^2}{2}-\frac{m_H^2 \eta^3}{2v}-\frac{m_H^2 \eta^4}{8v^2} \right] .
\end{equation}
Here $\eta$ is the real Higgs field - a physical degree of freedom after symmetry breaking. In this sense $\eta$ is a small oscillation around the minimum of the potential. The following must be true: $\eta >-v$, where $v$ is the vacuum expectation value of the Higgs field. The parameter $m_H$ is the bare Higgs mass before the volume renormalization. We go to the original form of the action by eliminating $\eta$ as follows
\begin{equation} \label{h}
h=v+ \eta
\end{equation}
where $h$ is the Higgs field before the small oscillation expansion. Using  Eq.~(\ref{h}) to eliminate $\eta$ from  Eq.~(\ref{S1}) we get
\begin{equation}
S=\int d^4x \left[ \frac{(\partial h) ^2}{2} - \frac{m_H^2 v^2}{8}+\frac{m_H^2 h^2}{4}-\frac{m_H^2 h^4}{8v^2} \right]
\end{equation}
$\eta>-v \rightarrow h>0$ and we follow the same procedure for the free field with the functional Schr$\ddot{\rm o}$dinger equation:
\begin{equation}\label{SE}
 -\frac{\partial_h^2 \Psi_h }{2 V} -\frac{V m_H^2 h^2}{4}\Psi_h+\frac{V m_H^2 h^4}{8v^2}\Psi_h= VH \Psi_h
\end{equation}
$\Psi_h$ is the wave-functional for the Higgs field $h$. Since $H$ represents the energy density, the eigenvalues $VH$ will have units of energy. The effective potential is
\be \label{veff}
V_{\rm eff}=-\frac{V m_H^2 h^2}{4}+\frac{V m_H^2 h^4}{8v^2}
 \ee
The physical non-perturbative mass of the Higgs particle will be defined as the energy difference between the ground and first excited state of the Higgs wave-functional for this effective potential. It is important to note that this non-pertubative mass is finite, and actually represents renormalized value of the bare mass $m_H$.

\begin{figure}[t]
   \centering
\includegraphics[width=3.5in]{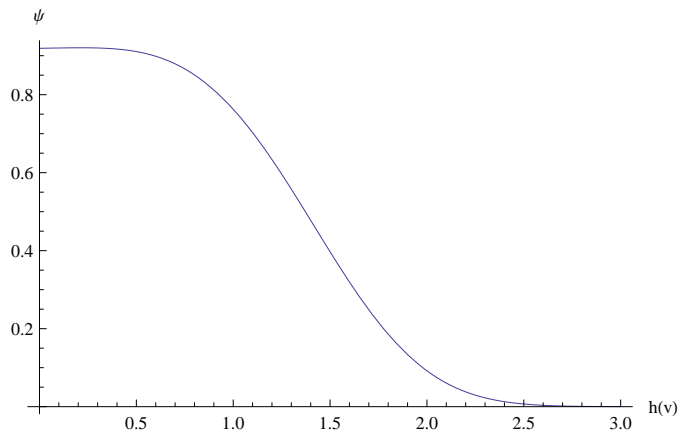}
\caption{The ground state wave-functional $\Psi_h$: $V^2H_0=0, \, (m_H^2 V^2)= 5.46755/v^4$. Near $h=v$, the amplitude of oscillations is not small.}
    \label{fig:wave1}
\end{figure}

We will treat Eq.~(\ref{SE}) non-perturbatively. We first impose the boundary condition $\partial_h \Psi_h |_{h=0}=0$ as it takes into account that the angular degrees of freedom are absorbed by the gauge fields. The second condition required is the normalization of the wave function. We numerically solve Eq.~(\ref{SE}) and find normalized solutions for the wave-functional $\Psi_h$ with the boundary condition above. We concentrate on solutions corresponding to the ground and first excited states (which are the eigenvalues of Eq.~(\ref{SE})). If we multiply Eq.~(\ref{SE}) by $V$, the only quantity that specifies the solution is $m_H^2 V^2$ measured in units of $v$. To preserve symmetry breaking, the ground state energy must satisfy $VH_0 \leq 0$ (or $V^2H_0 \leq 0$ after we multiply it by $V$). This condition translates into
\be
(m_H^2 V^2) \geq 5.46755/v^4
 \ee
The numerical solution for the ground state wave function $\Psi_h$ with $(m_H^2 V^2) = 5.46755/v^4$ is plotted in Fig.~\ref{fig:wave1}. It is clear that near $h=v$, the amplitude of oscillations is not small.  We can now calculate the vacuum expectation value of $h$ in this state as $<h>= \int \Psi_h^2 \, h \, dh$ and we find  $<h>=0.6425v$ corresponding to $\eta=-0.3575v$. This vacuum expectation value is not equal to the standard model value of $<h>=v$. Though in this case we have the ground state in the broken symmetry phase, the vacuum expectation value of the Higgs field is not what we expect under the small perturbation assumption. This invalidates the standard picture of the Higgs particle created in the broken phase. This indicates that we have to impose stricter conditions in order to come closer to the usual spontaneous symmetry breaking picture in the standard model Higgs mechanism. We will do so by considering the excited states.
\begin{figure}[t]
   \centering
\includegraphics[width=3.5in]{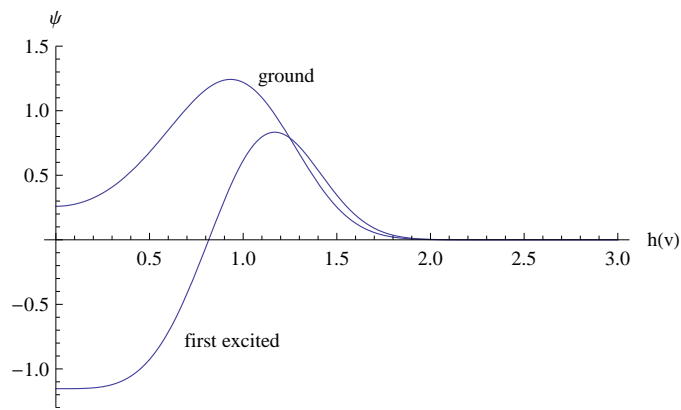}
\caption{The wave-functional $\Psi_h$ for the ground and first excited state, with the condition $(m_H^2 V^2)=82.7377/v^4$. The eigenvalues for the ground and first excited states are $V^2 H_0=-6.1474/v^2$ and $V^2H_1=0$ respectively. }
    \label{fig:wave2}
\end{figure}

\section{Excited States}

In order to have the Higgs particle created in the broken symmetry phase, the first excited state must satisfy $VH_1 \leq 0$ (or $V^2H_1 \leq 0$ after we multiply it by $V$), which is just the condition for the first excited state to be lower than the height of the barrier of the Higgs effective potential in Eq.~(\ref{veff}). This condition translates into
\be \label{con2}
(m_H^2 V^2) \geq 82.7377 /v^4.
\ee
Setting the minimal value that satisfies this condition $(m_H^2 V^2) = 82.7377/v^4$ we can plot the solutions for $\Psi_h$ in Fig.~\ref{fig:wave2}. This gives  $<h>=0.8815v$, which is closer to the standard model value. The  difference between the ground and first excited state is
\be
\Delta (V^2 H)=6.1474/v^2.
\ee
To get some explicit numerical estimates, we can fix the vacuum expectation value to $v=246$GeV.

At this point we can use the observed value of the physical Higgs mass of $M_H = 125$GeV. Since the physical Higgs mass is just the energy difference between the ground and first excited state, i.e. $M_H = \Delta (V^2 H)/V$, we can calculate the volume of the Higgs particle that saturates the bound in Eq.~(\ref{con2}) as
\be
V=\Delta (V^2 H)/M_H=\frac{6.1474}{v^2M_H} = 8.13 \times 10^{-7} GeV^{-3}
\ee
Therefore, Eq.~(\ref{con2}) implies that the physical volume of the Higgs particle must be larger than $V \sim 8.13\times 10^{-7}$GeV$^{-3}$.
We can use the same equation to find the value of the bare Higgs mass $m_H$ as
\be
m_H \geq \sqrt{82.7377} /v^2V = 184.88 GeV.
\ee
The fact that the physical Higgs mass is smaller than its bare mass, i.e. $M_H < m_H$, implies that the potential energy of the Higgs self-interaction reduces the physical mass, as it should.

Since Eq.~(\ref{con2}) is an inequality, one can always try choose values of parameters to get us closer to the standard model picture where $<h>=v$. We first note that it is impossible to get the strict equality $<h>=v$ with the given potential (\ref{veff}). This potential is not symmetric around $h=v$, so the solution for the wave function $\Psi_h$
will not be symmetric around $h=v$ (as can be seen in Fig.~\ref{fig:wave1} and Fig.~\ref{fig:wave2}). Then the expectation value $<h>= \int \Psi_h^2 \, h \, dh$ can not be strictly equal to $v$, unless the symmetry is somehow restored.

In addition, the freedom in Eq.~(\ref{con2}) is very limited from the physical requirement that the Compton wavelength must always be larger than the size of the particle. The Compton wavelength of a particle is an upper limit on the size of the particle - a Compton wavelength smaller then the physical size makes a particle ill defined. At distances smaller than the Compton wavelength, one cannot localize the particle - e.g. trying to localize an electron to within less than its Compton wavelength introduces uncertainties in its momentum so that it can have large enough energy to make an extra electron-positron pair. This will severely limit any freedom in adjusting the parameters in Eq.~(\ref{con2}).
Indeed, the value of the physical Higgs mass $M_H \sim 125$GeV gives the Compton wavelength of the Higgs particle, $L_C=\hbar /M_Hc$, which is very close to the physical size, $(3V/4\pi)^{1/3}$, of the particle. For the values calculated above we have
\be
(3V/4\pi)^{1/3} \geq 5.8 \times 10^{-3} GeV^{-1}
\ee
 and
\be
L_C=8 \times 10^{-3} GeV^{-1}
\ee
While the Compton wavelength is greater than the lower limit on the size of the particle, these two values are dangerously close to each other.

There are two competing effects in this context. Making the volume of the Higgs particle larger reduces the amplitude of oscillations (as demonstrated in Eq.~(\ref{pp})) and  justifies expansion around the vacuum. This will also get us closer to the usual perturbative picture of $<h>=v$. However, as the volume increases, the difference between the ground state and the first excited state becomes larger. This implies that the bare Higgs mass becomes smaller until finally reaching the limiting value of $m_H =M_H$. At that point Eq.~(\ref{SE}) reduces to that of a simple harmonic oscillator (the amplitude of oscillations, $h$, reduces and the quadratic term becomes dominant). However, as the volume increases, so does the size of the particle. Therefore, increasing the volume $V$ would make the physical Higgs particle just a usual small oscillation around the vacuum in a broken symmetry phase, however, this might make its size larger than the Compton wavelength and such a degree of freedom would not be a particle in the standard sense.

\section{Conclusion}

In conclusion, we took the first steps to formulate a non-perturbative treatment of QFT in which particles have finite size. We modified the commutation relation between the field and its canonical momentum to incorporate the finite size of the particle in a Lorentz invariant way. Using the examples of the free and self-interacting field theories we showed that divergences associated with point particles can be removed. In particular, a non-perturbative mass of a particle (the difference between the ground and first excited state) is finite. This non-perturbative approach is physically transparent and allowed us to address
the question that perviously went unchecked in the literature - whether a field can be condensed enough to be observed as a well
behaved particle. By treating particles as finite size objects, we have one more control parameter at our disposal - the Compton wavelength of the
particle that we are trying to observe. A well defined particle must have its Compton wavelength larger than the physical size of the particle. In the case of the spontaneous symmetry breaking, e.g. Higgs mechanism, the Higgs particle has to be created in the broken symmetry phase, or otherwise the produced degree of freedom will not have the properties we expect. Using the value of $M_H=125$GeV, we showed that the range of parameters which allows for a well defined Higgs particle is very restricted.  This method can likely be extended to all interacting field theories. For example including fermions and gauge bosons in the effective potential in Eq.~(\ref{SE}) would complete the analysis of the Higgs field and strengthen the conclusions drawn here.

Since the volume is included in the quantization, every interaction/vertex will be non-local, which in ceratin extent may modify phenomenology. We emphasis that this non-locality only implies that interactions do not happen at the point, but are actually smeared out. The correct thing to do is to actually use the wave packet of the particle which will give us the correct wavefunction distribution in the interaction terms. The non-local effects will then be included through the commutator in Eq.~(\ref{commutation}). This is along the lines of the standard quantization procedure where the commutator tells the Lagrangian that the interactions happen at one single point.

\begin{acknowledgments}
The authors thank A. Weltman, M. Abbott,  M. Seikel and  Robert de Mello Koch for helpful discussions and correspondence. DS acknowledges the financial support from NSF, grant number PHY-1066278. DCD acknowledges the financial support from Shanghai Institutions of Higher Learning, the Science and Technology Commission of Shanghai Municipality,  grant No.11DZ2260700.
\end{acknowledgments}

\end{document}